\documentclass[letterpaper,10pt,conference]{ieeeconf}

\usepackage{cite}
\usepackage{graphicx}
\usepackage{tikz}
\usepackage{subcaption}
\usepackage[cmex10]{amsmath}
\usepackage{algpseudocode}
\usepackage{algorithmicx} 
\usepackage{array}
\usepackage{url}
\usepackage{mathrsfs}

\usepackage{color}
\usepackage{multirow}
\usepackage{cite}
\usepackage{balance}
\usepackage{multicol,lipsum}

\usepackage{color}
\usepackage{multirow}
\usepackage{cite}
\usepackage{balance}
\usepackage{amssymb}
\usepackage{verbatim}
\usepackage{cuted}
\usepackage{flushend}

\usepackage{epsfig}
\usepackage{amsfonts,amssymb,multirow,bigstrut,booktabs,ctable,latexsym}

\usepackage[mathscr]{eucal}
\usepackage{verbatim}

\usepackage{amsthm}
\usepackage{algorithm}

\usepackage{stmaryrd}

\IEEEoverridecommandlockouts  
\title{\Large\bf Small-Signal Dynamics of Lossy Inverter-Based Microgrids for Generalized Droop Controls}

	\author{Abdullah Al Maruf$^{1}$, Anamika Dubey$^{2}$, Sandip Roy$^{3}$
     \thanks{$^1$California State University, Los Angeles.
        {\tt\small amaruf@calstatela.edu}}
		\thanks{$^{2}$ Washington State University. 
			{\tt\small anamika.dubey@wsu.edu}}
        \thanks{$^{3}$ Texas A\&M University.
        {\tt\small sandip@tamu.edu}}
		\thanks{}
	}

\begin{document}
\maketitle

\begin{abstract}

A network-level small-signal model is developed for lossy microgrids, which considers coupled angle and voltage dynamics of inverter-based microgrids and uses a more general framework of droop controls in the inverter. It is shown that when relative resistances of the lines in the microgrid are reasonably consistent and differences of voltage angles across the lines are small at the operating point, the generalized droop controls can be designed to enforce decoupling between angle dynamics and voltage dynamics. Next, structural results for the asymptotic stability of small-signal angle and voltage dynamics are given for the case when generalized droop control achieves decoupling. Simulated transient responses of a modified IEEE 9-bus system are presented to validate the theoretical findings which show the effectiveness of generalized droop controls in independently shaping the settling times of the angle and voltage responses of the lossy microgrid system. 

\end{abstract}

\section{Introduction}

A microgrid is an interconnected low/medium-voltage power distribution network primarily supplied by inverter-based distributed energy resources (DERs) \cite{bouzid2015survey, fazal2023grid}. Droop controls are popularly used in inverter-based microgrids to achieve power-sharing between multiple DERs by mimicking the inertial dynamics and governor controls of generators in the bulk grid. Droop control does not require any communication to operate, which makes it more appealing to use \cite{chandorkar1993control}. Traditionally, two separate droop controls, i.e., $P-\omega$ and $Q-V$ droop controls, are used in the microgrid to regulate the voltage angles and voltage magnitudes of the grid based on the standard decoupling assumption \cite{chandorkar1993control,guerrero2010hierarchical, eberlein2021small}. While this decoupling assumption simplifies analysis (which mimics the analysis for the bulk grid) as undertaken in \cite{egwebe2016implementation,song2015small,al2019small,al2021small}, it is often not accurate for lossy microgrids since lines in power distribution networks have substantially large resistance to inductance ratio and also may exhibit voltage magnitude/angle differentials across them \cite{guerrero2010hierarchical,vasquez2009adaptive}. This poses a significant concern when using traditional droop-controls in microgrids to achieve stability and the desired power sharing.

In this study, we consider a generalization of the conventional droop controller where the grid-side voltage frequency and magnitude are both set in a co-dependent way based on the locally measured real and reactive power \cite{guerrero2010hierarchical,vasquez2009adaptive}. This {\em generalized droop} scheme was proposed with the idea that it can approximately decouple the angle (i.e., voltage angle) and voltage (i.e., voltage magnitude) dynamics in lossy microgrids, and hence should allow systematic shaping of these two dynamics and power sharing. However, a network-level model and analysis of microgrids with generalized droop have yet to be undertaken in the literature. Furthermore, it is necessary to develop structural analyses of the microgrid dynamics -- i,e., ones that are phrased in terms of the graph topology of the microgrid and other general characteristics of the system rather than the specific parameter values because of high variability in operating conditions for microgrids.  We point out that the important study of Schiffer et al \cite{schiffer2014conditions} captures the coupling of the voltage and angle dynamics for traditional droops, but it does not obtain structural characterizations of stability or design rubrics for the lossy case.

One main focus of this study is to develop an accurate model and structural analysis for lossy microgrids that use generalized droop controls. Our main contributions in this study are the following:

1)  A complete signal-signal model for an islanded microgrid's network-level dynamics is developed, which captures resistive losses, generalized droop controls, smoothing/filtering used in droop controls, and multiple bus types.




2)  A design of generalized droop is presented, which can approximately decouple the voltage magnitude and angle dynamics and thus achieve desirable small-signal properties, provided that the $R/X$ ratios of lines are reasonably consistent across the microgrid and bus angle differences across the lines are small at the microgrid's operating point.


3) For the droop control that achieves the decoupling, structural small-signal stability results are presented. These results mainly follow from our previous studies \cite{al2019small,al2021small}, but these studies pre-supposed decoupling and did not consider the generalized droop controls.  

4) Simulated small-disturbance responses of a modified IEEE 9-bus system are presented to verify our results. We specifically show how the settling time of small disturbance responses of voltage angles and magnitudes can be shaped by appropriately designing generalized droop controls.


\section{Small-Signal model for A Lossy Microgrid with Generalized Droop Controls}
\label{Chap11:sec:Modeling}
The transient and small-signal dynamics of a lossy microgrid operating in islanded mode are modeled in this section. The model presented here incorporates 1) a generalized droop scheme for the primary controller at inverter-based generators, 2) low-pass filtering of measured local powers implemented at the droop controls, 3) heterogeneous (i.e., inverter and non-inverter) bus types, and 4) resistive losses in network lines.  First, a nonlinear differential-algebraic-equation (DAE) model is developed. Then, two simplifications -- an approximation of the algebraic equations as dynamic ones via singular perturbation, and linearization -- are undertaken to obtain a network-level small-signal model. The derived small-signal model is suitable for the analysis and control of small-signal properties of microgrids. 

\subsection{Microgrid Network: Nonlinear DAE Model}
\label{Chap11:subsec:Microgrid Network: Nonlinear DAE Model Formulation}

A nonlinear DAE model is developed for a lossy microgrid that uses generalized droop controls at inverter-based generators, which also generalizes previously-developed network-level models for conventional droop-controlled microgrids \cite{al2019small,al2021small,schiffer2014conditions}. We present the microgrid network model in three steps: 1) the model for the generalized droop control (Section \ref{Chap11:subsec:Droop-Controlled Inverter Model}); 2) the network power-flow model (Section \ref{Chap11:subsec:Network Model}); and 3) the full DAE model which combines the network power-flow models with dynamic models of the DERs (Section \ref{Chap11:subsec:DAE Model for the Microgrid Network}).

\subsubsection{Droop-Controlled Inverter Model}
\label{Chap11:subsec:Droop-Controlled Inverter Model}
A grid-forming inverter operating in an islanded microgrid acts as an ideal AC voltage source, where its voltage (both magnitude and frequency) can be assigned at will by the fast inner control systems. To support power sharing among multiple inverters, conventional droop controls are usually used as a primary controller, which mimics the governor control of synchronous generators in the bulk grid in the sense that a $P-\omega$ droop control regulates frequency (i.e., frequency of voltage) based on the measured active power while in a similar way a $Q-V$ droop control regulates voltage (i.e. voltage magnitude) based on the measured reactive power at the connected bus. These independent controls of frequency and voltage are ineffective for lossy microgrids as the decoupling assumption between angle dynamics and voltage does not hold. A generalization of conventional droop controls has been proposed to achieve better performance for lossy microgrids, where both frequency and voltage are set in a co-dependent way based on the measured real and reactive powers \cite{guerrero2010hierarchical}. This generalized control offers an additional design parameter to modify the governing equations for setting the frequency and voltage of an inverter. In this study, we consider the generalized droop control scheme in the inverters and develop a network-level model of a microgrid that employs such a scheme.

Droop controls are incorporated with low pass filters to suppress the high-frequency variations in measured powers. Here, in our framework,  the filter and the time delay due to fast inner control loops are abstracted as a first-order low pass filter. The model of an inverter at bus $i$ with a generalized droop control scheme is depicted in Fig. \ref{Chap11:fig1}. Mathematically, for the generalized droop control, the frequency and voltage at the inverter can be written in terms of the real and reactive power in the Laplace domain as follows:

\begin{figure}[t]
\centering
\includegraphics[scale=0.3]{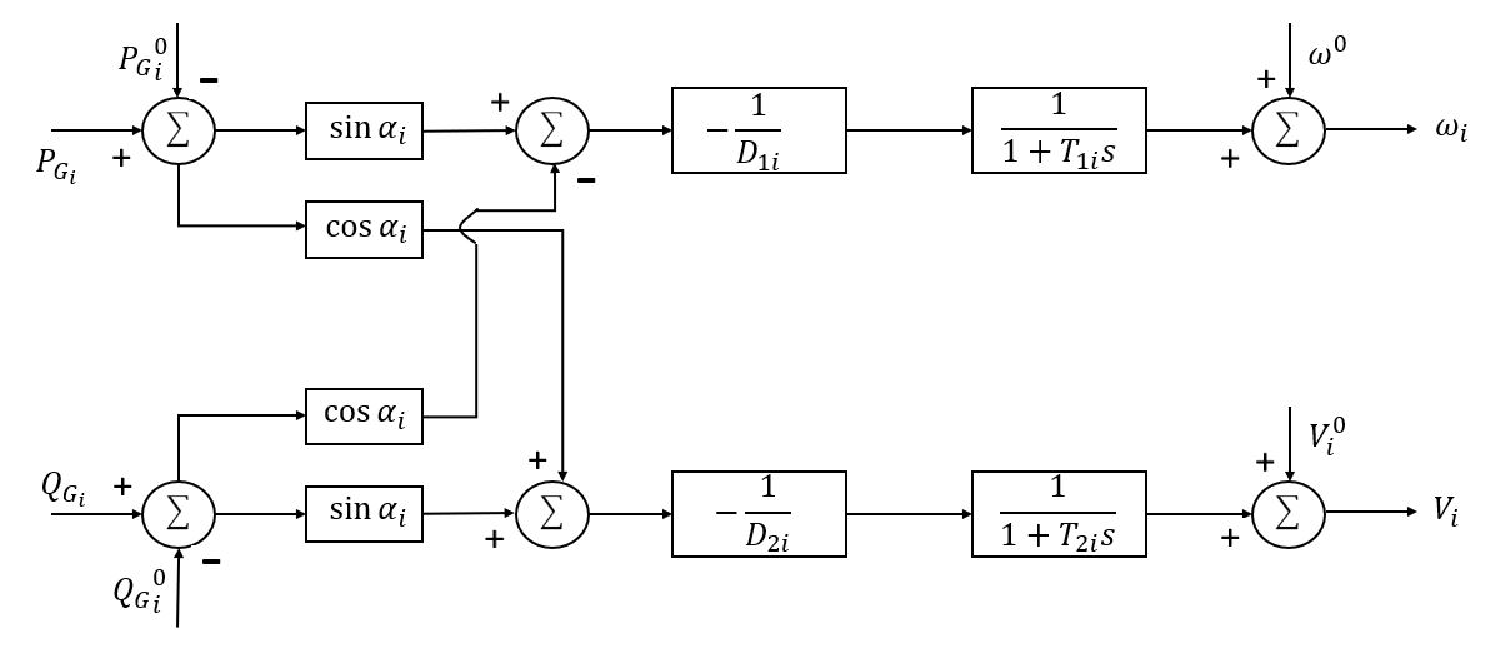}
\caption{Block diagram of generalized droop control with filtering.}
\label{Chap11:fig1}
\end{figure}
	
\begin{eqnarray}\label{Chap11:eq1}
\big((P_{G_i}(s)-P^0_{G_i}) \sin(\alpha_i) -(Q_{G_i}(s)-Q^0_{G_i}) \cos(\alpha_i)\big)  \nonumber \\
\times \frac{-1/D_{1i}}{1+T_{1i}s}=\omega_i(s)-\omega^0
\end{eqnarray}
and,
\begin{eqnarray}\label{Chap11:eq1b}
\big((P_{G_i}(s)-P^0_{G_i}) \cos(\alpha_i) +(Q_{G_i}(s)-Q^0_{G_i}) \sin(\alpha_i)\big)  \nonumber \\
\times \frac{-1/D_{2i}}{1+T_{2i}s}=V_i(s)-V_i^0
\end{eqnarray}
\noindent where, $\omega^0$ is the common nominal angular frequency\footnote{$\omega^0$ is ideally set to $0$ since droop control operates relative to the nominal rotational frequency e.g. 377 rad/s in North America.}; $\omega_i(s)$ is the Laplace transform of frequency $\omega_i$; $V_i^0$ is the nominal voltage; $V_i(s)$ is the Laplace transform of voltage $V_i$;  $P^0_{G_i}$ and $Q^0_{G_i}$ are the reference active power and reactive power respectively; $P_{G_i}(s)$ and $Q_{G_i}(s)$ are the Laplace transform of generated active power and generated reactive power respectively; $D_{1i}, D_{2i}\ge 0$ are the reciprocals of droop gains of the inverter; $T_{1i}, T_{2i}$ are the time constants of the low pass filters and $\alpha_i$ is an additional design parameter in this droop law which corresponds to an angle. This angle specifies the co-dependence between two droop equations and is introduced to reduce the coupling between the angle and voltage dynamics. We note that the traditional droop law is recovered when this angle is selected as $\alpha_i= \pi/2$ in (\ref{Chap11:eq1}) and (\ref{Chap11:eq1b}). Equations (\ref{Chap11:eq1}) and (\ref{Chap11:eq1b}) can be rewritten in the time domain as:
\begin{eqnarray}\label{Chap11:eq2}
(P_{G_i}-P^0_{G_i}) \sin(\alpha_i) -(Q_{G_i}-Q^0_{G_i}) \cos(\alpha_i) \nonumber \\ =D_{1i}\omega^0-D_{1i}\dot{{\theta }_i}-D_{1i}T_{1i}\ddot{{\theta }_i}
\end{eqnarray}
and,
\begin{eqnarray}\label{Chap11:eq2b}
(P_{G_i}-P^0_{G_i}) \cos(\alpha_i) +(Q_{G_i}-Q^0_{G_i}) \sin(\alpha_i)   \nonumber \\
=D_{2i}V_i^0-D_{2i}{V_i}-D_{2i}T_{2i}\dot{V_i}
\end{eqnarray}

Equations (\ref{Chap11:eq2}) and (\ref{Chap11:eq2b}) govern the voltage and angle dynamics of inverter-based buses with generalized-droop control.

\subsubsection{Network Model}
\label{Chap11:subsec:Network Model}

Here we derive the standard network model for power flow, which includes lossy or resistive lines. We consider a connected microgrid network with $n$ buses, labeled $i=1,\hdots, n$. The admittance of the line between buses $i$ and $k$ obtained from the bus-admittance matrix is denoted by $Y_{ik}\angle \phi_{ik}= G_{ik}+jB_{ik}$ where $G_{ik}$ and $B_{ik}$ are respectively conductance and susceptance of the line according to the bus-admittance matrix. Note for regular power system, $\pi /2 \leq \phi_{ik} \leq \pi$ for $i \neq k$. The active and reactive power flow equations for bus $i$ are given as the following \cite{grainger1999power}:
\begin{eqnarray}\label{Chap11:eq3}
P_{G_i} = P_{L_i} &+& V_i^2 G_{ii} \nonumber \\
&+& \sum_{k \in adj(i)}{V_i V_k Y_{ik}~\cos (\theta_i-\theta _k- \phi_{ik})} 
\end{eqnarray}
and,
\begin{eqnarray}\label{Chap11:eq3b}
Q_{G_i} = Q_{L_i} &-& V_i^2 B_{ii} \nonumber \\
&+& \sum_{k \in adj(i)}{V_i V_k Y_{ik}~\sin (\theta_i-\theta _k- \phi_{ik})} 
\end{eqnarray}
where, $P_{G_i}$ and $P_{L_i}$ are the active power generation and demand at bus $i$, similarly $Q_{G_i}$ and $Q_{L_i}$ are the reactive power generation and demand at bus $i$. Here $adj(i)$ refers to the set of buses that are adjacent to bus $i$ (or equivalently, the set of buses connected to bus $i$ via a distribution line). 

In this study, we consider the constant power model for the loads, which is commonly used in the stability analysis of bulk-grid and microgrid \cite{grainger1999power,schiffer2014conditions}. According to the constant power load model, the active and reactive power for the load at bus $i$ can be expressed as: $P_{L_i}=P^0_{L_i}$ and $Q_{L_i}=Q^0_{L_i}$ where both $P^0_{L_i}$ and $Q^0_{L_i}$ are arbitrary constants.

\subsubsection{DAE Model for the Microgrid Network}
\label{Chap11:subsec:DAE Model for the Microgrid Network}

The DAE model for the microgrid is developed by integrating the network model and the generalized-droop-control model. Let us first denote the set of buses that have inverters as $\mathcal{V}_A$. Using (\ref{Chap11:eq2}), (\ref{Chap11:eq2b}), (\ref{Chap11:eq3}), (\ref{Chap11:eq3b}) and the load model, the governing differential equations for the bus $i \in \mathcal{V}_A$ can be written as:
\begin{align} \label{Chap11:eq4}
\small
&D_{1i}T_{1i}\ddot{{\theta}_i} =\big((P^0_{G_i}-P^0_{L_i})\sin(\alpha_i)- (Q^0_{G_i}-Q^0_{L_i})\cos(\alpha_i) \nonumber \\
&~~~~+D_{1i}\omega^0\big)-D_{1i}\dot{\theta}_i -V_i^2\big(G_{ii}\sin(\alpha_i)+B_{ii}\cos(\alpha_i)\big) \nonumber \\&~~~-\sum_{k\in adj (i)}{V_iV_k{Y}_{ik}~\cos(\theta_i-\theta _k- \phi_{ik}+\pi/2-\alpha_i)}
\end{align}
and,
\begin{align} \label{Chap11:eq4b}
\small
&D_{2i}T_{2i}\dot{V_i}=\big((P^0_{G_i}-P^0_{L_i})\cos(\alpha_i)+ (Q^0_{G_i}-Q^0_{L_i})\sin(\alpha_i) \nonumber \\
&~~~~+D_{2i}V_i^0\big)-D_{2i}V_i+V_i^2(-G_{ii}\cos(\alpha_i)+B_{ii}\sin(\alpha_i)) \nonumber \\ 
&~~~~-\sum_{k\in adj (i)}{V_iV_k{Y}_{ik}~\sin (\theta_i-\theta _k- \phi_{ik}+\pi/2-\alpha_i)}
\end{align}

We next present the governing equations for the buses that do not have inverter-based generators $(P_{G_i}=0, ~Q_{G_i}=0)$. It is convenient to denote the set of these buses as $\mathcal{V}_B$. Using (\ref{Chap11:eq3}), (\ref{Chap11:eq3b}) and the load model, the governing equations for the bus $i\in \mathcal{V}_B$ can be written as:

\begin{equation} \label{Chap11:eq5new}
\small
0=-P^0_{L_i} -V_i^2G_{ii} -\sum_{k\in adj (i)}{V_iV_k{Y}_{ik}~\cos (\theta_i-\theta _k- \phi_{ik})}
\end{equation}
and,
\begin{equation} \label{Chap11:eq5newb}
\small
0=-Q^0_{L_i}+V_i^2B_{ii} -\sum_{k\in adj (i)}{V_iV_k{Y}_{ik}~\sin (\theta_i-\theta _k - \phi_{ik})}
\end{equation}

Equations (\ref{Chap11:eq4}), (\ref{Chap11:eq4b}), (\ref{Chap11:eq5new}) and (\ref{Chap11:eq5newb}) in together describe nonlinear DAE model of microgrid.

\subsection{Simplified Models for Analysis}
\label{Chap11:subsec:Simplified Modeling for Analysis}
We develop a linear differential-equation approximation of the nonlinear DAE to enable formal analysis of small-signal properties. First, a singular perturbation argument is applied to obtain a differential-equation approximation of the DAE in a way that maintains the topological structure of the network (hence, we refer to this as a structure-preserving model). Then, the nonlinear differential-equation model is linearized to obtain the small-signal model of the microgrid dynamics.

\subsubsection{Structure-Preserving Model}
\label{Chap11:subsec:Structure-Preserving Model}
Here we apply a singular perturbation argument to approximate the algebraic equations as fast dynamics yielding a differential-equation model. The singular-perturbation approach has the advantage of preserving the topological structure of the microgrid, which is helpful for structural analysis of the microgrid. Before applying singular perturbation approximation, we undertake a transformation on the set of algebraic equations of the DAE to match expressions of the differential equations. This will allow us to obtain a compact and suitable expression for the eventual model, while sufficing that the dynamical properties of the original DAE model are maintained in the approximated differential equation model (with the undertaken transformation). 

In analogy with the generalized droop controls of inverter-based bus, for each bus $i \in \mathcal{V}_B$ we consider an angle $\alpha_i$ where $0\leq\alpha_i\leq\pi/2$ to undertake the following transformation on (\ref{Chap11:eq5new}) and (\ref{Chap11:eq5newb}).

\begin{eqnarray} \label{Chap11:eq5}
\small
0=(-P^0_{L_i}\sin(\alpha_i)+Q^0_{L_i}\cos(\alpha_i)) \nonumber \\ -V_i^2\big(G_{ii}\sin(\alpha_i)+B_{ii}\cos(\alpha_i)\big) \nonumber \\ -\sum_{k\in adj (i)}{V_iV_k{Y}_{ik}~\cos(\theta_i-\theta _k- \phi_{ik}+\pi/2-\alpha_i)}
\end{eqnarray}
and,
\begin{eqnarray} \label{Chap11:eq5b}
\small
0=(-P^0_{L_i}\cos(\alpha_i)-Q^0_{L_i}\sin(\alpha_i)) \nonumber \\
+V_i^2(-G_{ii}\cos(\alpha_i)+B_{ii}\sin(\alpha_i)) \nonumber \\ -\sum_{k\in adj (i)}{V_iV_k{Y}_{ik}~\sin (\theta_i-\theta _k- \phi_{ik}+\pi/2-\alpha_i)}
\end{eqnarray}

Now we apply singular perturbation approximations directly on (\ref{Chap11:eq5}) and (\ref{Chap11:eq5b}). Upon appropriate selection of $\alpha_i$, based on the singular-perturbation argument we can say that each $i\in {\mathcal{V}}_B$ bus approximately has the following dynamics:

\begin{eqnarray} \label{Chap11:eq6}
\small
\epsilon_1\ddot{\theta_i}=(-P^0_{L_i}\sin(\alpha_i)+Q^0_{L_i}\cos(\alpha_i)) -\epsilon_2 {\dot{\theta }}_i \nonumber \\ -V_i^2\big(G_{ii}\sin(\alpha_i)+B_{ii}\cos(\alpha_i)\big) \nonumber \\ -\sum_{k\in adj (i)}{V_iV_k{Y}_{ik}~\cos(\theta_i-\theta _k- \phi_{ik}+\pi/2-\alpha_i)}  
\end{eqnarray}
and,
\begin{eqnarray} \label{Chap11:eq6b}
\small
\epsilon_3\dot{V_i}=(-P^0_{L_i}\cos(\alpha_i)-Q^0_{L_i}\sin(\alpha_i)) \nonumber \\ +V_i^2(-G_{ii}\cos(\alpha_i)+B_{ii}\sin(\alpha_i)) \nonumber \\ -\sum_{k\in adj (i)}{V_iV_k{Y}_{ik}~\sin (\theta_i-\theta _k- \phi_{ik}+\pi/2-\alpha_i)}
\end{eqnarray}
\noindent where $\epsilon_1, \epsilon_2$ and $\epsilon_3$ are sufficiently small positive numbers. Some discussions on these factors are worthwhile. First, note that the approximated differential equation system and the original DAE system have the same equilibrium solutions. Specifically in our case, the system given by (\ref{Chap11:eq4}), (\ref{Chap11:eq4b}), (\ref{Chap11:eq5new}), (\ref{Chap11:eq5newb}) and the system given by (\ref{Chap11:eq4}), (\ref{Chap11:eq4b}), (\ref{Chap11:eq6}), (\ref{Chap11:eq6b}) have the same equilibrium solutions. Additionally, provided that the factors $\epsilon_1$, $\epsilon_2$, $\epsilon_3$ are chosen appropriately (specifically, $\epsilon_2$ and $\epsilon_3$ are sufficiently small and $\epsilon_1$ is on the order of $\epsilon_2^2$) and noting that both systems depend on the relative angles of bus voltage rather than their actual values (thus, a common zero eigenvalue is maintained by the Jacobians of both systems at equilibrium), the local stability of the equilibrium of approximate model (\ref{Chap11:eq4}), (\ref{Chap11:eq4b}), (\ref{Chap11:eq6}), (\ref{Chap11:eq6b}) suffices the local stability of the DAE model (\ref{Chap11:eq4}), (\ref{Chap11:eq4b}), (\ref{Chap11:eq5new}), (\ref{Chap11:eq5newb}) according to the singular perturbation theory \cite{kokotovic1999singular,chang1995direct}. Therefore, we will use differential equation set (\ref{Chap11:eq4}), (\ref{Chap11:eq4b}), (\ref{Chap11:eq6}), (\ref{Chap11:eq6b}) for deriving the structural stability conditions given that the above factors are selected appropriately. The appropriate bounds for these factors can be derived from the singular-perturbation literature \cite{kokotovic1999singular}. 


To enable structural analysis for the stability of microgrid, it is convenient to present a compact form of the approximated model based on a graph-theoretic description of the microgrid. First, we consider a directed graph $\mathcal{G}$ defined for the microgrid, where a vertex represents each bus, and the line between buses $i$ and $k$ corresponds to two directed edges $(i,k)$ and $(k,i)$. Hence, the vertex set $\mathcal{V}$ has cardinality $n$ and edge set $\mathcal{E}\subseteq \mathcal{V}\times\mathcal{V}$ has cardinality $2l$, where $l$ is the number of lines in the microgrid. For convenience, we arbitrarily label and order the edges, with edge $e_m$ ($m=1,\hdots, 2l$) representing edge $(i,k)$ of the graph. Then, the incidence matrix $\boldsymbol{E} \in \mathbb{R}^{n\times 2l}$ of $\mathcal{G}$ is defined as  $E_{im}=1$ and $E_{km}=-1$ for each $e_m \in \mathcal{E}$, with all other entries being zero. Furthermore, the orientation matrix is defined for the directed graph (see \cite{goldin2013weight}) as a matrix $\boldsymbol{C}\in \mathbb{R}^{n\times 2l}$ with entries $C_{im}=1$ if $E_{im}=1$ and zero otherwise. We also use the following notations: $\boldsymbol{x}=[x_i]$ denotes a vector $\boldsymbol{x}=[x_1,x_2, \cdots ,x_n]^T \in \mathbb{R}^n$, and $diag(\boldsymbol{x})$ denotes a diagonal matrix in $\mathbb{R}^{n\times n}$ whose diagonal entries are $x_1,x_2, \cdots ,x_n$ respectively. Using these notations, we can write (\ref{Chap11:eq4}), (\ref{Chap11:eq4b}), (\ref{Chap11:eq6}), (\ref{Chap11:eq6b}) as vector equations as follows.
\begin{eqnarray} \label{Chap11:eq7}
\boldsymbol{M}_P\ddot{\boldsymbol{\theta }}=\boldsymbol{P}^0-\boldsymbol{D}_P\dot{\boldsymbol{\theta}}-(diag(\boldsymbol{V}))^2 \hat{\boldsymbol{G}} \nonumber \\-\boldsymbol{C} \boldsymbol{U} \cos(\boldsymbol{E}^T \boldsymbol{\theta }-\boldsymbol{\phi}+\frac{\pi}{2}\boldsymbol{1}_n-\boldsymbol{\alpha})
\end{eqnarray}
\begin{eqnarray} \label{Chap11:eq7b}
\boldsymbol{M}_Q\dot{\boldsymbol{V}}=\boldsymbol{Q}^0-\boldsymbol{D}_Q\boldsymbol{V}+(diag(\boldsymbol{V}))^2 \hat{\boldsymbol{B}} \nonumber \\ 
-\boldsymbol{C} \boldsymbol{U} \sin(\boldsymbol{E}^T \boldsymbol{\theta }-\boldsymbol{\phi}+\frac{\pi}{2}\boldsymbol{1}_n-\boldsymbol{\alpha})
\end{eqnarray}

In above the vectors/matrices represent corresponding physical quantities or parameters of the microgrid. Specifically,  $\boldsymbol{\theta} = [\theta_i] \in \mathbb{R}^n$ and $\boldsymbol{V} = [V_i] \in \mathbb{R}^n$ denote the vectors of bus voltage phase angles and magnitude respectively. $\boldsymbol{\alpha} = [\alpha_i] \in \mathbb{R}^n$ whereas $\boldsymbol{\phi}=[\phi_{ik}] \in \mathbb{R}^{2l}$ is the vector containing phase angles of the admittance of each directed lines.  $\boldsymbol{M}_p~\in \mathbb{R}^{n\times n}$ is a diagonal matrix imitating inertia, where $M_{p_{ii}}=D_{1i}T_{2i}$ if $i\in \mathcal{V}_A$, otherwise $M_{P_{ii}}= \epsilon_1$. Similarly, $\boldsymbol{M}_Q~\in \mathbb{R}^{n\times n}$ is also a diagonal matrix, where $M_{Q_{ii}}=D_{2i}T_{2i}$ if $i\in \mathcal{V}_A$, otherwise $M_{Q_{ii}}= \epsilon_3$. $\boldsymbol{P}^0= [P^0_i] ~\in\mathbb{R}^n$ denotes nominal injected (transformed) active power in each bus, where $P^0_i=(P^0_{G_i}-P^0_{L_i})\sin(\alpha_i)- (Q^0_{G_i}-Q^0_{L_i})\cos(\alpha_i)$
$+D_{1i}\omega^0$ if $i\in \mathcal{V}_A$, otherwise $P^0_i=(-P^0_{L_i}\sin(\alpha_i)+Q^0_{L_i}\cos(\alpha_i))$. In the same manner, $\boldsymbol{Q}^0= [Q_i^0] ~\in\mathbb{R}^n$ denotes nominal (transformed) injected reactive power in each bus, where $Q^0_i=(P^0_{G_i}-P^0_{L_i})\cos(\alpha_i)+ (Q^0_{G_i}-Q^0_{L_i})\sin(\alpha_i)+D_{2i}V_i^0$ if $i\in \mathcal{V}_A$, otherwise $Q^0_i=-P^0_{L_i}\cos(\alpha_i)-Q^0_{L_i}\sin(\alpha_i)$. $\boldsymbol{D}_P~\in \mathbb{R}^{n \times n}$ is a diagonal matrix representing damping coefficients, where $D_{P_{ii}}=D_{1i}$ if $i\in \mathcal{V}_A$, otherwise $D_{P_{ii}}=\epsilon_2$. Similarly, $\boldsymbol{D}_Q~\in \mathbb{R}^{n \times n}$ is also a diagonal matrix, where $D_{Q_{ii}}=\ D_{2i}$ if $i\in \mathcal{V}_A$, otherwise $D_{Q_{ii}}=0$. $\boldsymbol{U} \in \mathbb{R}^{2l \times 2l}=diag(V_iV_kY_{ik})$ is a diagonal matrix in above whose diagonal entries represent the magnitudes of active/reactive powers flow through the edges. $\hat{\boldsymbol{G}}$ and $\hat{\boldsymbol{B}}$ are vectors in $\mathbb{R}^n$ where  $\hat{G}_{i}=G_{ii}\sin(\alpha_i)+B_{ii}\cos(\alpha_i)$ and $\hat{B}_{i}=-G_{ii}\cos(\alpha_i)+B_{ii}\sin(\alpha_i)$ respectively. $\boldsymbol{1}_n$ denotes a vector in $\mathbb{R}^n$ with all entry equal to 1. It should be noted that in (\ref{Chap11:eq7}) and (\ref{Chap11:eq7b}) the functions $\cos(.)$ and $\sin(.)$ are meant in the Hadamard sense (i.e. component-wise). 

Next, we use equations (\ref{Chap11:eq7}) and (\ref{Chap11:eq7b}) to develop the small-signal model.


\subsubsection{Small-Signal Model}
\label{Chap11:subsec:Small-Signal Model}

Linearization is undertaken around an arbitrary equilibrium to obtain a small-signal differential-equation model for the microgrid network, which allows the study of small signal properties at an equilibrium of interest. Now, we point out an interesting characteristic of the equilibrium of microgrid systems. We note that each equilibrium of a microgrid system corresponds to a manifold since the microgrid's dynamics only depend on the differences of phase angles of bus voltages (see the model given by (\ref{Chap11:eq7}) and (\ref{Chap11:eq7b}) or equivalently (\ref{Chap11:eq4}), (\ref{Chap11:eq4b}), (\ref{Chap11:eq5}), (\ref{Chap11:eq5b})). Therefore, if ${\boldsymbol{x}^s=\begin{bmatrix} \boldsymbol{\theta}^s & \boldsymbol{\dot{\theta}}^s & \mathbf{V}^s \end{bmatrix}}^T$ is an equilibrium of a microgrid system, then $\boldsymbol{x}^s +k \boldsymbol{v}_0$ is also an equilibrium of the system (or vice versa), where ${\boldsymbol{v}_0=\begin{bmatrix} \boldsymbol{1}_n & {\bf 0}_n & {\bf 0}_n \end{bmatrix}}^T$ (note, $\boldsymbol{0}_n$ is a vector in $\mathbb{R}^n$ with all entries equal to $0$) and $k \in \mathbb{R}$. From the power system viewpoint, these are physically the same equilibrium as only the relative phase angles among the buses are of interest. Therefore, from this point, by an equilibrium $\boldsymbol{x}^s$ of a microgrid system, we refer to the manifold $\boldsymbol{x}^s +k \boldsymbol{v}_0$. Now, let us define the state vector as ${\boldsymbol{x}=\begin{bmatrix} \boldsymbol{\theta} & \boldsymbol{\dot{\theta}}& \mathbf{V} \end{bmatrix}}^T$ and denote an equilibrium point (or the manifold) as $\boldsymbol{x}^s=\begin{bmatrix} \boldsymbol{\theta}^s & \boldsymbol{\dot{\theta}}^s & \mathbf{V}^s \end{bmatrix}^T$.
By linearizing (\ref{Chap11:eq7}) and (\ref{Chap11:eq7b}) around the equilibrium point $\boldsymbol{x}^s$ (or more accurately around the equilibrium manifold $\boldsymbol{x}^s +k \boldsymbol{v}_0$ as discussed before) we get:
\vspace{-0.2cm}
\begin{equation} \label{Chap11:eq8}
\boldsymbol{\Delta }\dot{\boldsymbol{x}}=\begin{bmatrix} \boldsymbol{\Delta \dot{\theta}} \\ \boldsymbol{ \Delta \ddot{\theta}} \\ \boldsymbol{\Delta \dot{V}} \end{bmatrix}=\boldsymbol{J}(\boldsymbol{x}^s) \begin{bmatrix} \boldsymbol{\Delta \theta} \\ \boldsymbol{\Delta \dot{\theta}} \\ \boldsymbol{\Delta V} \end{bmatrix}=\boldsymbol{J}(\boldsymbol{x}^s)\boldsymbol{\Delta x}
\end{equation}

The Jacobian matrix $\boldsymbol{J}\left({\boldsymbol{x}}^s\right)$ of the nonlinear model evaluated at the equilibrium is given in the appendix. In this Jacobian, $\boldsymbol{W}_1(\boldsymbol{x}^s)=\frac{\partial  (\boldsymbol{U}^s \cos(\boldsymbol{E}^T \boldsymbol{\theta^s }-\boldsymbol{\phi}+\frac{\pi}{2}\boldsymbol{1}_n-\boldsymbol{\alpha}))}{\partial (\boldsymbol{E}^T \boldsymbol{\theta^s }-\boldsymbol{\phi}+\frac{\pi}{2}\boldsymbol{1}_n-\boldsymbol{\alpha})}= diag \big(-\boldsymbol{U}^s \sin (\boldsymbol{E}^T \boldsymbol{\theta^s }-\boldsymbol{\phi}+\frac{\pi}{2}\boldsymbol{1}_n-\boldsymbol{\alpha}) \big)$ $\in \mathbb{R}^{2l \times 2l}$ and $\boldsymbol{W}_2(\boldsymbol{x}^s)=\frac{\partial (\boldsymbol{U}^s \sin(\boldsymbol{E}^T \boldsymbol{\theta^s }-\boldsymbol{\phi}+\frac{\pi}{2}\boldsymbol{1}_n-\boldsymbol{\alpha}))}{\partial (\boldsymbol{E}^T \boldsymbol{\theta^s }-\boldsymbol{\phi}+\frac{\pi}{2}\boldsymbol{1}_n-\boldsymbol{\alpha})}= diag \big(\boldsymbol{U}^s \cos(\boldsymbol{E}^T \boldsymbol{\theta^s }-\boldsymbol{\phi}+\frac{\pi}{2}\boldsymbol{1}_n-\boldsymbol{\alpha}) \big)$ $\in \mathbb{R}^{2l \times 2l}$. Here, $\boldsymbol{0}_{n \times n}$ and $\boldsymbol{I}_{n \times n}$ denote $n \times n$ zero matrix and $n \times n$ unit matrix respectively. 

Here we emphasize that the linearization undertaken in (\ref{Chap11:eq7}) and (\ref{Chap11:eq7b}) should be understood as a linearization over the manifold $\boldsymbol{x}^s +k \boldsymbol{v}_0$. Therefore the linearized system holds an invariant manifold where all the angles are synchronized and all the frequency and voltage deviations are zero. In analogy with \cite{song2015small},  we define $\boldsymbol{W}_1(\boldsymbol{x}^s)$ and $\boldsymbol{W}_2(\boldsymbol{x}^s)$ as the edge weights of two graphs $\mathcal{G}_1$ and graph $\mathcal{G}_2$ respectively. This definition allows us to rewrite the dynamics in terms of the (directed) Laplacian matrix of these graphs, which is helpful for the analysis of the stability. Following the existing literature \cite{song2015small}, we consider the weighted directed graphs $\mathcal{G}_1(\boldsymbol{x}^s)=(\mathcal{V} ,\mathcal{E},\boldsymbol{W}_1(\boldsymbol{x}^s))$ and $\mathcal{G}_2(\boldsymbol{x}^s)=(\mathcal{V} ,\mathcal{E},\boldsymbol{W}_2(\boldsymbol{x}^s))$. We note that the edge weights in the graphs $\mathcal{G}_1(\boldsymbol{x}^s)$ and $\mathcal{G}_2(\boldsymbol{x}^s)$ are different and depend on the operating point $\boldsymbol{x}^s$ and the parameter $\boldsymbol{\alpha}$, but the vertex and edge sets of both graphs are same as graph $\mathcal{G}$. The Laplacian matrix for these directed weighted graphs can be written as $\boldsymbol{L}(\mathcal{G}_1(\boldsymbol{x}^s))=\boldsymbol{C}\boldsymbol{W}_1 (\boldsymbol{x}^s) {\boldsymbol{E}}^{{T}}$ and $\boldsymbol{L}(\mathcal{G}_2(\boldsymbol{x}^s))=\boldsymbol{C}\boldsymbol{W}_2 (\boldsymbol{x}^s) {\boldsymbol{E}}^{{T}}$. Therefore, we can simplify the Jacobian as:

\begin{strip}
\begin{eqnarray} \label{Chap11:eq11}
\scriptsize
\boldsymbol{J}(\boldsymbol{x}^s)=\begin{bmatrix}
 \boldsymbol{0}_{n\times n} & \boldsymbol{I}_{n\times n} & \boldsymbol{0}_{n\times n}\\
-\boldsymbol{M}_P^{-1}\boldsymbol{L}(\mathcal{G}_1(\boldsymbol{x}^s)) & -\boldsymbol{M}_P^{-1}\boldsymbol{D}_P & -\boldsymbol{M}_P^{-1} \Big(- \boldsymbol{L}(\mathcal{G}_2(\boldsymbol{x}^s)) +2~diag(\hat{\boldsymbol{P}}^s) \Big) (diag (\boldsymbol{V}^s))^{-1} \\
-\boldsymbol{M}_Q^{-1}\boldsymbol{L}(\mathcal{G}_2(\boldsymbol{x}^s)) & \boldsymbol{0}_{n\times n} & -\boldsymbol{M}_Q^{-1} \Big( \boldsymbol{L}(\mathcal{G}_1(\boldsymbol{x}^s)) +2~  diag(\hat{\boldsymbol{Q}}^s)  +\boldsymbol{D}_Q~diag(\boldsymbol{V}^s) \Big) (diag (\boldsymbol{V}^s))^{-1}
\end{bmatrix}  
\end{eqnarray} 
\end{strip}

Here $\hat{\boldsymbol{P}}^s=diag(\hat{\boldsymbol{G}})  (diag(\boldsymbol{V}^s))^2+ diag(\boldsymbol{C} \boldsymbol{U}^s$ $\cos(\boldsymbol{E}^T \boldsymbol{\theta^s }-\boldsymbol{\phi}+\frac{\pi}{2}\boldsymbol{1}_n-\boldsymbol{\alpha}))$ and $\hat{\boldsymbol{Q}}^s=-diag(\boldsymbol{\hat{B}}) (diag(\boldsymbol{V}^s))^2$ $+ diag(\boldsymbol{C} \boldsymbol{U}^s$ $\sin(\boldsymbol{E}^T \boldsymbol{\theta^s }-\boldsymbol{\phi}+\frac{\pi}{2}\boldsymbol{1}_n-\boldsymbol{\alpha})$ denote the vectors whose entries are related to injected powers of the buses at equilibrium. Specifically, $\hat{P}_i^s=(V_i^s)^2\big(G_{ii}\sin(\alpha_i)$ $+B_{ii}\cos(\alpha_i)\big)$ $+\sum_{k\in adj (i)}{V_i^sV_k^s{Y}_{ik}~\cos(\theta_i^s-\theta _k^s- \phi_{ik}+\pi/2-\alpha_i)}$ and $\hat{Q}_i^s= (V_i^s)^2(G_{ii}\cos(\alpha_i)$ $-B_{ii}\sin(\alpha_i))$  $+\sum_{k\in adj (i)}{V_i^sV_k^s{Y}_{ik}~\sin (\theta_i^s-\theta _k^s- \phi_{ik}+\pi/2-\alpha_i)}$.

Note that the parameter $\alpha_i$ for $i\in {\mathcal{V}}_B$ is arbitrary given that it is appropriately chosen following the singular perturbation argument. Also note that by substituting $\alpha= \pi/2$ in (\ref{Chap11:eq11}), we obtain the small-signal model of the microgrid for the traditional $P-\omega$ and $Q-V$ droop control. Therefore, (\ref{Chap11:eq11}) provides a more general form for the small-signal model of a droop-controlled inverter-based microgrid. Next, we present a specific droop control design that can enforce decoupling between angle and voltage dynamics.

\section{Design of Generalized Droop Control to Achieve Decoupling}
\label{Chap11:SSSA}

In this section,  we discuss the design of generalized droop controls so that decoupling can be achieved between angle and voltage dynamics. Specifically, we want to select the parameter $\alpha_i$, which will enforce decoupling in dynamics in a lossy microgrid. Recall that $\alpha_i$ represents a physically meaningful angle for inverter buses; for non-inverter buses, $\alpha_i$ is an arbitrary angle which merely introduces a mathematical transformation. 

We select the parameter $\alpha_i$ leveraging some practical facts. As distribution lines of a typical microgrid are usually manufactured from same constituent materials, $R/X$ ratios of the lines lie in a close range of values. Based on this fact it is reasonable to assume that $\phi_{ik}$ of each line $(i,k)$ is approximately equal to a fixed value $\phi_0$ i.e. $\phi_{ik} \approx \phi_0$ for each $(i,k) \in \mathcal{E}$. (Note, impedance angle $\beta_{ik}=\tan^{-1}(R_{ik}/X_{ik})$ is related to the admittance angle in Y-bus matrix as: $\beta_{ik}=\pi-\phi_{ik}$.) Additionally, it is not too restrictive to assume that the differences of bus angles across the lines at equilibrium (i.e. $|\theta_i^s-\theta_k^s|$) are small enough. This is a standard assumption in bulk grid which is typically true for microgrid when the operating point of the inverters is around the rated condition. Besides, microgrids typically have radial structures and the differences of bus angles at equilibrium are known to be small for radial networks. Under these assumptions here we show that it is possible to select the parameter $\alpha_i$ in a way so that decoupling can be achieved. Specifically, we show that setting $\alpha_i=\pi-\phi_0=\beta_0$ for each bus $i$ leads to decoupling where $\beta_0$ is the impedance angle of all the lines (i.e. $\phi_{ik} = \phi_0$ for all $(i,k) \in \mathcal{E}$).

First note that by substituting $\phi_{ik} = \phi_0$ and $\alpha_i=\pi-\phi_0$, the edge weights $\boldsymbol{W}_1(\boldsymbol{x}^s))$ and $\boldsymbol{W}_2(\boldsymbol{x}^s))$ of the graphs $\mathcal{G}_1(\boldsymbol{x}^s)$ and $\mathcal{G}_2(\boldsymbol{x}^s)$ can be written as:  $\boldsymbol{W}_1(\boldsymbol{x}^s))= diag \big(\boldsymbol{U}^s \cos(\boldsymbol{E}^T \boldsymbol{\theta}^s) \big)$ and $\boldsymbol{W}_2(\boldsymbol{x}^s))= diag \big(\boldsymbol{U}^s \sin (\boldsymbol{E}^T \boldsymbol{\theta}^s) \big)$. Note that the edge weights of both $(i,k)$ and $(k,i)$ edges in $\mathcal{G}_1(\boldsymbol{x}^s)$ become equal to $V_i^s V_k^s Y_{ik}\cos(\theta_i^s -\theta_k^s)$. Hence $\mathcal{G}_1(\boldsymbol{x}^s)$ can be considered as an undirected graph where each line between bus $i$ and $k$ corresponds to an undirected edge $(i,k)$. For undirected graph, the Laplacian matrix is symmetric and thereby $\boldsymbol{L}(\mathcal{G}_1(\boldsymbol{x}^s))$ can be written as $\boldsymbol{L}(\mathcal{G}_1(\boldsymbol{x}^s))=\boldsymbol{E}_u \boldsymbol{W}_u(\boldsymbol{x}^s) \boldsymbol{E}_u^{T}$ where $\boldsymbol{E}_u \in \mathbb{R}^{n\times l}$ is the incidence matrix and $\boldsymbol{W}_u(\boldsymbol{x}^s)= diag(V_i^sV_k^sY_{ik}) \cos {(\boldsymbol{E}_u^T \boldsymbol{\theta}^s)} \in \mathbb{R}^{l \times l}$ is the weights of the edges in the undirected graph $\mathcal{G}_1(\boldsymbol{x}^s)$. Following the existing literature \cite{song2015small,song2017network}, we refer this weighted undirected graph $\mathcal{G}_1(\boldsymbol{x}^s)$ as {\it active power flow graph}. Now, applying small angle difference assumption, $\boldsymbol{L}(\mathcal{G}_2(\boldsymbol{x}^s))$ can be approximated as zero since $\boldsymbol{W}_2(\boldsymbol{x}^s))= diag \big(\boldsymbol{U}^s \sin (\boldsymbol{E}^T \boldsymbol{\theta}^s) \big) \approx \mathbf{0}$. Furthermore, noting that $\frac {B_{ii}}{G_{ii}}=\tan(\phi_0)$, we find $\hat{\boldsymbol{P}}^s \approx \mathbf{0}$ as $\hat{P}_i^s= (V_i^s)^2\big(G_{ii}\sin(\alpha_i)$ $+B_{ii}\cos(\alpha_i)\big)$ $+\sum_{k\in adj (i)}{V_i^sV_k^s{Y}_{ik}~\sin(\theta_i^s-\theta _k^s)}$. By substituting these results we can approximate eqn. (\ref{Chap11:eq11}) as eqn. (\ref{Chap11:eq12}).

From the block form of (\ref{Chap11:eq12}) it is obvious that when the above assumptions hold and $\alpha_i$ in each bus is chosen as $\alpha_i=\pi-\phi_0=\beta_0$, the angle and voltage dynamics become decoupled. Thus, by selecting $\alpha_i$ in each generalized droop control the same as the impedance angle of the lines, we can achieve decoupling. This result, in fact, has been recognized for a single-bus system \cite{guerrero2010hierarchical} but has not been formally analyzed for a network consisting of multiple inverters. Thus, we have extended and verified the decoupling-achieving design of generalized droop controls for a network. Note, for non-inverter buses $\alpha_i$ is arbitrary and thus we can free to select $\alpha_i=\pi-\phi_0=\beta_0$ for non-inverter buses also in our model.
 
\noindent {\bf Remark:} When the above assumptions (i.e. impedance angle is constant throughout the network and bus angle differences are sufficiently small) do not hold, then the edge weights $\boldsymbol{W}_2(\boldsymbol{x}^s))$ can not be approximated to zero. As a consequence, we can not conclude decoupling between the angle and voltage dynamics for the proposed selection of $\alpha_i$. In that case, the norm of $\boldsymbol{W}_2(\boldsymbol{x}^s)$ provides us a measure of coupling between angle and voltage dynamics of microgrid. However, in practice these assumptions are not required to be satisfied strictly. Specifically, when $R/X$ ratios of lines lie within a small range and bus angle differences are relatively small, we can use an average of the $R/X$ ratios in the design of generalized droop control to achieve decoupling in an approximate sense and thereby predict stability or shape small signal response reasonably (see simulation examples). 

\begin{strip}
\begin{eqnarray} \label{Chap11:eq12}
\scriptsize
\boldsymbol{J}(\boldsymbol{x}^s) = \begin{bmatrix}
 \boldsymbol{0}_{n\times n} & \boldsymbol{I}_{n\times n} & \boldsymbol{0}_{n\times n}\\
-\boldsymbol{M}_P^{-1}\boldsymbol{L}(\mathcal{G}_1(\boldsymbol{x}^s)) & -\boldsymbol{M}_P^{-1}\boldsymbol{D}_P & \boldsymbol{0}_{n\times n}\\
\boldsymbol{0}_{n\times n} & \boldsymbol{0}_{n\times n} & -\boldsymbol{M}_Q^{-1} \Big( \boldsymbol{L}(\mathcal{G}_1(\boldsymbol{x}^s)) +2~  diag(\hat{\boldsymbol{Q}}^s)  +\boldsymbol{D}_Q~diag(\boldsymbol{V}^s) \Big) (diag (\boldsymbol{V}^s))^{-1}
\end{bmatrix}  
\end{eqnarray} 
\end{strip}

\section{Small-Disturbance Analysis under decoupling-Achieving Design}

The decoupling achieving droop control simplifies the analysis of the model significantly as the analysis of angle and voltage dynamics can be undertaken separately in an independent manner. It also enables us to shape the small disturbance responses of a lossy microgrid effectively. To begin with our analysis, we make the following assumptions to guarantee that the model \eqref{Chap11:eq12} validly represents the small signal model of a lossy microgrid for the selected generalized droop control.

\noindent {\bf Assumptions 1:} 
\begin{enumerate}
    \item Phase angles differences (i.e. $|\theta_i^s-\theta_k^s|$) across the lines are sufficiently small.
    \item The impedance angles (or equivalently $R/X$ ratios) of all lines are approximately same.
    \item The parameter $\alpha_i$ is chosen as $\alpha_i=\pi-\phi_0=\beta_0$ for each inverter bus where $\beta_0$ is the common impedance angle of the lines.  
\end{enumerate}

The above assumptions ensure that the selected droop control achieves decoupling and \eqref{Chap11:eq12} represents a valid small-signal model. Now we provide separate structural analyses for the angle dynamics and voltage dynamics of the model \eqref{Chap11:eq12} with a particular focus on stability. 

We begin by noting that from the block form of (\ref{Chap11:eq12}) it is obvious that the small-signal angle dynamics of the microgrid is dictated by the following state matrix:
\begin{eqnarray} \label{Chap11:eq13}
\boldsymbol{J}_A(\boldsymbol{x}^s)=\begin{bmatrix}
 \boldsymbol{0}_{n\times n} & \boldsymbol{I}_{n\times n}\\
-\boldsymbol{M}_P^{-1}\boldsymbol{L}(\mathcal{G}_1(\boldsymbol{x}^s)) & -\boldsymbol{M}_P^{-1}\boldsymbol{D}_P
\end{bmatrix}  
\end{eqnarray}

These dynamics are identical to the small signal angle dynamics derived for conventional $P-\omega$ droop under the decoupling assumption  \cite{al2019small}. In \cite{al2019small}, we showed that when the Laplacian matrix $\boldsymbol{L}(\mathcal{G}_1(\boldsymbol{x}^s))$ satisfies the conditions --  i.e., the matrix is positive semidefinite and has a non-repeated eigenvalue at zero, the angle dynamics is asymptotically stable irrespective of filtering constant or droop gain. The Laplacian is guaranteed to be positive semi-definite and has a non-repeated eigenvalue at zero, if its off-diagonal entries are nonnegative and the matrix is irreducible (equivalently, its associated graph is connected and has positive edge weights). For the Laplacian matrix in our formulation, the weight of the edge between vertices $i$ and $k$ is given by  $V_i^s V_k^s Y_{ik}\cos{(\theta_i^s -\theta_k^s)}$.  This weight is positive if the difference between the power-flow bus angle across the corresponding line is less than $\pi/2$, or more generally $|\theta_i^0-\theta_k^0|~ mod ~2\pi < \pi/2$. Since this condition is automatically satisfied in our case due to the small angle difference assumption, the small signal model (\ref{Chap11:eq13}) is readily asymptotically stable. These observations are summarized in the following theorem (presented without proof since this is an automatic consequence of the results derived in \cite{al2019small}).

\medskip

\noindent \textbf{Theorem 1:}  \textit{Consider the microgrid model governed by (\ref{Chap11:eq4}), (\ref{Chap11:eq4b}), (\ref{Chap11:eq5new}) and (\ref{Chap11:eq5newb}). Given the conditions mentioned in Assumptions 1 are satisfied, the small-signal angle dynamics of the microgrid (\ref{Chap11:eq13}) is asymptotically stable for any positive droop gain and filtering time constant.}
\medskip


Now we we turn our attention to the analysis of voltage dynamics of small-signal model given by (\ref{Chap11:eq11}). First note that from the block form of (\ref{Chap11:eq11}) that the small-signal voltage dynamics of the microgrid is dictated by the following state matrix:
\begin{eqnarray} \label{Chap11:eq15}
\small
\boldsymbol{J}_V(\boldsymbol{x}^s)=-\boldsymbol{M}_Q^{-1} \Big( \boldsymbol{L}(\mathcal{G}_1(\boldsymbol{x}^s)) +2~  diag(\hat{\boldsymbol{Q}}^s)  \nonumber \\ +\boldsymbol{D}_Q~diag(\boldsymbol{V}^s) \Big) 
(diag (\boldsymbol{V}^s))^{-1}
\end{eqnarray}

These dynamics are also identical to the small signal voltage dynamics derived for conventional $Q-V$ droop under the decoupling assumption \cite{al2021small}. Following the same approach in \cite{al2021small},  $\boldsymbol{L}(\mathcal{G}_1(\boldsymbol{x}^s)) +2~  diag(\hat{\boldsymbol{Q}}^s)$ can be seen as the Laplacian of an enhanced version of the graph $\mathcal{G}_1(\boldsymbol{x}^s)$ where each vertex has a self-loop with weight $2\hat{Q}_i^s$ $=2(V_i^s)^2(G_{ii}\cos(\alpha_i)$ $-B_{ii}\sin(\alpha_i))$  $-2\sum_{k\in adj (i)}{V_i^sV_k^s{Y}_{ik}~\cos (\theta_i^s-\theta _k^s)}$. Thereby for the purpose of structural insights here we introduce a self-looped version of active power flow graph $\mathcal{G}_{lp}(\mathbf{x}^s)$  $=(\mathcal{V} ,\mathcal{E},\mathbf{W}_u(\mathbf{x}^s), diag(2 \hat{\mathbf{Q}}^s))$ which we call \textit{loopy active power flow graph}. Therefore we can write $\mathbf{L}(\mathcal{G}_{lp}(\mathbf{x}^s))= \mathbf{L}(\mathcal{G}_1(\mathbf{x}^s)) + diag(2 \mathbf{Q}^s) = \boldsymbol{E}_u \boldsymbol{W}_u(\boldsymbol{x}^s) \boldsymbol{E}_u^{T} + 2~  diag(\mathbf{Q}^s) $. Substituting this in (\ref{Chap11:eq15}) we get:
\begin{equation} \label{Chap11:eq16}
\small \mathbf{J}_V(\mathbf{x}^s)= -\mathbf{M}_Q^{-1} \big( \mathbf{L}(\mathcal{G}_{lp}(\mathbf{x}^s)) +\mathbf{D}_Q~diag(\mathbf{V}^s)\big) (diag (\mathbf{V}^s))^{-1} 
\end{equation}

Now we can readily apply the graph-theoretic stability conditions for the small-signal model (\ref{Chap11:eq16}) from our previous study. Following Theorem 1 of \cite{al2021small} we can state the following result, which gives conditions on the Laplacian of the loopy active power flow graph under which the voltage dynamics are asymptotically stable for any droop gains and filter time constants (presented without proof since this is an automatic consequence of the results derived in \cite{al2021small}).

\medskip

\noindent \textbf{Theorem 2}  \textit{Consider the microgrid model governed by (\ref{Chap11:eq4}), (\ref{Chap11:eq4b}), (\ref{Chap11:eq5new}) and (\ref{Chap11:eq5newb}). Given the conditions mentioned in Assumptions 1 are satisfied, the small-signal voltage dynamics of the microgrid (\ref{Chap11:eq16}) is asymptotically stable for any positive droop gain and filtering time constant if the Laplacian $\mathbf{L}(\mathcal{G}_{lp}(\mathbf{x}^s))$ is positive definite.}

\medskip

We refer readers to \cite{al2021small} for further discussion on structural results on voltage stability. It is important to note that the derived stability conditions in the Theorems 1 and 2 are independent of droop and filtering parameters. Although these conditions do not depend on the droop gain $D$ and filtering time constant $T$, these parameters can affect other performance metrics such as power sharing and settling time of the angle and voltage response. Since the developed stability conditions are independent of droop and filtering parameters, our results allow the design of these parameters at will to achieve other desired performance goals (e.g., power sharing and settling time) while ensuring stability.

\section{Simulation Example}

Here we simulate transient response of an example inverter-based microgrid employing generalized droop controls. For this, we use a modified IEEE 9-bus system (adapted to capture characteristics of a typical microgrid). The network diagram of the modified IEEE 9-bus network and model parameters for this example are detailed in \cite{webpage} due to space constraint. We note that the modified microgrid network is made to be radial (the line between Bus 5 and Bus 6 was removed from the IEEE 9-bus network) to make angle differences at equilibrium small. We consider lossy lines where $R/X$ for each line follows a normal distribution with mean $0.7$ and standard deviation $0.02$. The network is operating at equilibrium A as detailed in \cite{webpage}. The maximum angle difference across lines at this equilibrium is obtained as $3.2265^0$. For the generalized droop, we select $\alpha_i=\tan^{-1}(0.7)$ as suggested in our analysis. We consider a small disturbance to the bus voltages away from the equilibrium A and simulate the small disturbance response of the microgrid for different filter time constants. 

\begin{figure}[thpb]
\centering
\includegraphics[width=9cm, height=8cm]{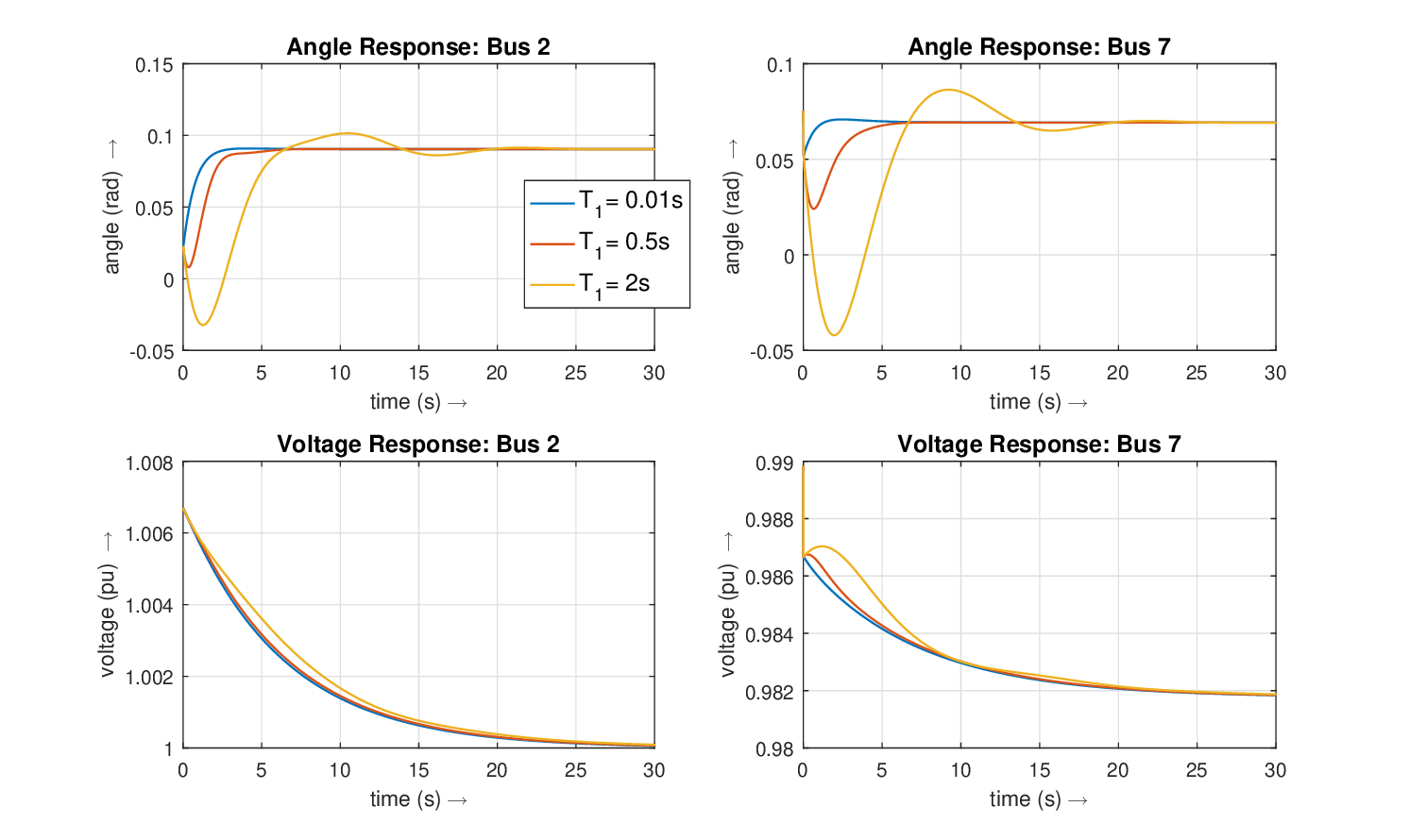}
\caption{Small-disturbance angle and voltage response of modified IEEE 9-bus lossy system with radial topology for generalized droop with $T_{1i} =0.01, 0.5$ and $2$ sec in the droop controls.}
\label{Chap11:fig3}
\end{figure} 

\begin{figure}[thpb]
\centering
\includegraphics[width=9cm, height=8cm]{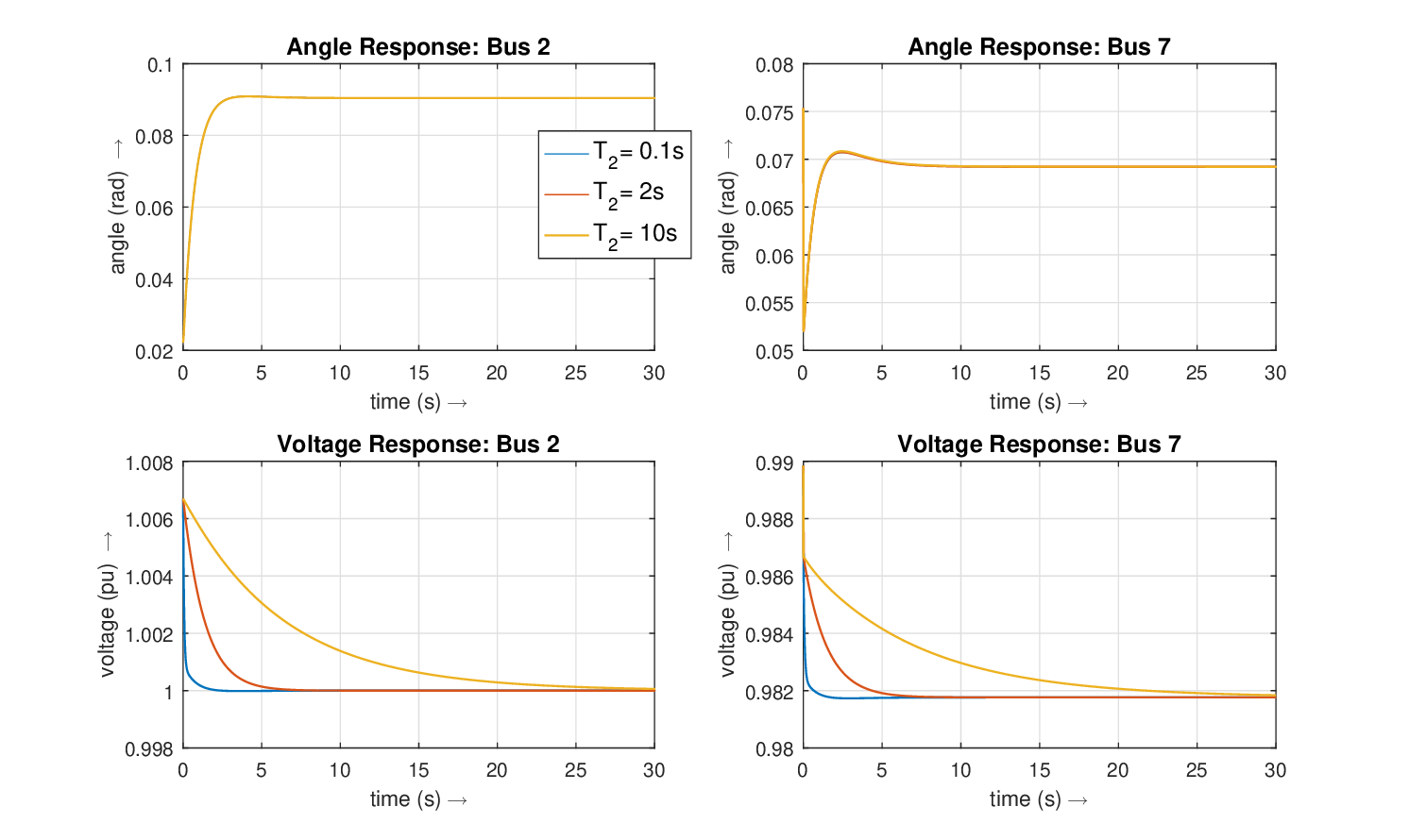}
\caption{Small-disturbance angle and voltage response of modified IEEE 9-bus lossy system with radial topology for generalized droop with $T_{2i} =0.1, 2$ and $10$ sec in the droop controls.} \label{Chap11:fig4}
\end{figure} 

Since bus voltage angles are small and $\alpha_i=\tan^{-1}(0.7)$,  Theorem 1 implies that the small-signal angle dynamics of the system is asymptotically stable regardless of the filter time constant. We also note that the Laplacian matrix $\mathbf{L}(\mathcal{G}_{lp})$ is positive definite. Therefore, according to Theorem 2, the small-signal voltage dynamics of the system is asymptotically stable regardless of the filter time constant. Figures \ref{Chap11:fig3} and \ref{Chap11:fig4} show the angle and voltage response of an inverter bus (Bus 2) and non-inverter bus (Bus 7) for different time filter constants. We notice that for all these time constants, both angle and voltage dynamics are asymptotically stable, as implied by the theorems.

Now we turn our attention to the settling times of the responses. In  Fig. \ref{Chap11:fig3}, $T_{1i}$ is varied to $0.01$s, $0.5$s and $2$s whereas $T_{2i}$ is kept fixed to $10$s in the generalized droop controls of the inverters. Similarly, in Fig. \ref{Chap11:fig4}, $T_{2i}$ is varied to $0.1$s, $2$s and $10$s whereas $T_{1i}$ is kept fixed to $0.01$s in the generalized droop controls of the inverters. In Fig. \ref{Chap11:fig3}, we see that settling times of angle responses are increased with the increase of filtering time constant $T_{1i}$ whereas voltage responses remain almost the same. Similarly, in Fig. \ref{Chap11:fig4} we see that settling times of voltage responses are increased with the increase of filtering time constant $T_{2i}$ whereas angle responses remain exactly the same. This example suggests that the designed generalized droop control enforces decoupling between angle and voltage dynamics effectively and allows us to independently shape settling times of both angle and voltage responses via selecting appropriate filtering time constants $T_{1i}$ and $T_{2i}$.

\section{Conclusions}
A network-level structure preserving small-signal model has been developed for generalized droop control. It has been verified that for arbitrary but relatively fixed $R/X$ ratios of lines decoupling can still be achieved by appropriate selection of droop law. For the decoupling-achieving design, several structural conditions have been presented for the asymptotic stability of both angle and voltage regardless of the filtering and droop parameters of controls. Simulation results have been presented that support the derived theoretical results and show generalized droop control's ability to shape small disturbance responses of microgrid using our model. We stress that the presented research represents only an initial step toward a comprehensive stability analysis of microgrids; key next steps include analysis for more general microgrid models with variable $R/X$ ratio, and application of the analyses in the design of larger systems.

\section{Appendix}

\begin{strip}
\begin{eqnarray} \label{Chap11:eq9}
\tiny
\begin{bmatrix}
\boldsymbol{0}_{n\times n} & \boldsymbol{I}_{n\times n} & \boldsymbol{0}_{n\times n}\\
-\boldsymbol{M}_P^{-1}\boldsymbol{C}\boldsymbol{W}_1(\boldsymbol{x}^s)\boldsymbol{E}^T & -\boldsymbol{M}_P^{-1}\boldsymbol{D}_P & -\boldsymbol{M}_P^{-1} \Big(- \boldsymbol{C}\boldsymbol{W}_2(\boldsymbol{x}^s)\boldsymbol{E}^T +2~ diag(\hat{\boldsymbol{G}})  (diag(\boldsymbol{V}^s))^2+ 2~diag(\boldsymbol{C} \boldsymbol{U}^s \cos(\boldsymbol{E}^T \boldsymbol{\theta^s }-\boldsymbol{\phi}+\frac{\pi}{2}\boldsymbol{1}_n-\boldsymbol{\alpha})) \Big) (diag (\boldsymbol{V}^s))^{-1} \\
-\boldsymbol{M}_Q^{-1}\boldsymbol{C}\boldsymbol{W}_2(\boldsymbol{x}^s)\boldsymbol{E}^T & \boldsymbol{0}_{n\times n} & -\boldsymbol{M}_Q^{-1} \Big( \boldsymbol{C}\boldsymbol{W}_1(\boldsymbol{x}^s)\boldsymbol{E}^T -2~  diag(\boldsymbol{\hat{B}}) (diag(\boldsymbol{V}^s))^2+ 2~diag(\boldsymbol{C} \boldsymbol{U}^s \sin(\boldsymbol{E}^T \boldsymbol{\theta^s }-\boldsymbol{\phi}+\frac{\pi}{2}\boldsymbol{1}_n-\boldsymbol{\alpha})) +\boldsymbol{D}_Q~diag(\boldsymbol{V}^s)\Big) (diag (\boldsymbol{V}^s))^{-1}  
\end{bmatrix} 
\nonumber
\end{eqnarray} 
\end{strip}


\begin{figure}[!] 
\centering
\includegraphics[scale=0.3]{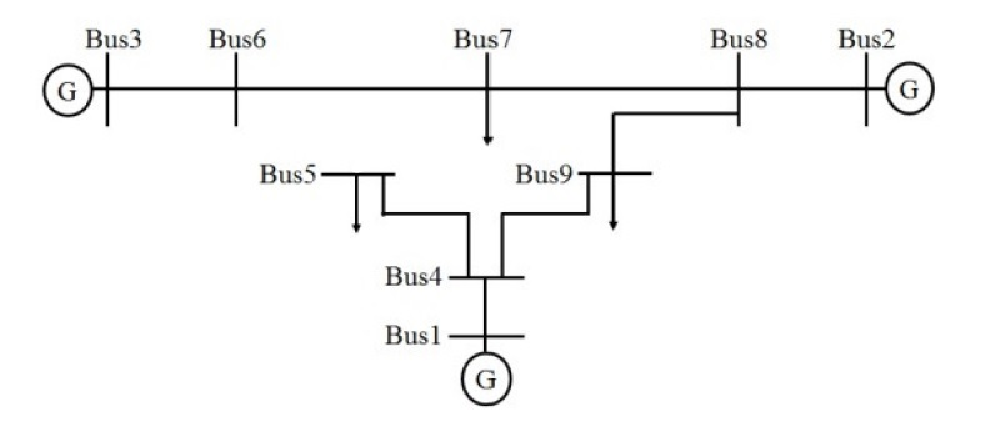}
\caption{Diagram of the modified IEEE 9-bus lossy test system.}
\label{fig:1}
\end{figure}

\begin{table}[!]
\centering
\caption{Bus parameters of modified IEEE 9-bus lossy test system}
\begin{tabular}{|l|l|l|l|l|}
\hline
Bus & $P_{G_i}^0-P_{L_i}^0$ (p.u.) & $Q_{G_i}^0-Q_{L_i}^0$ (p.u.) & $D_{1i}$ & $D_{2i}$ \\ \hline
1 &   &   & 5 & 10 \\ \hline
2 & 0.3260 &   & 5 &  10\\ \hline
3 & 0.1700 &   & 5 &  10 \\ \hline
4 & 0 &  0  & $\epsilon_2$  & 0 \\ \hline
5 & -0.18 &  -0.12  & $\epsilon_2$  & 0 \\ \hline
6 & 0 &  0   & $\epsilon_2$   & 0 \\ \hline
7 & -0.2 &  -0.4  & $\epsilon_2$  & 0 \\ \hline
8 & 0 &  0  & $\epsilon_2$  & 0 \\ \hline
9 & -0.25 &  -0.6  & $\epsilon_2$  & 0 \\ \hline
\end{tabular}
\end{table}

\begin{table}[!]
\centering
\caption{Line parameters of modified IEEE 9-bus lossy test system}
\begin{tabular}{|l|l|l|l|l|}
\hline
Line & $R_i$ (p.u.) & $X_i$ (p.u.) \\ \hline
  (1,4)& 0.0387   & 0.0576 \\ \hline
  (4,5)& 0.0648   & 0.0920 \\ \hline
  (3,6)& 0.0412   & 0.0586 \\ \hline
  (6,7)& 0.0703   & 0.1008 \\ \hline
  (7,8)& 0.0517   & 0.0720 \\ \hline
  (8,2)& 0.0433   & 0.0625 \\ \hline
  (8,9)& 0.1100   & 0.1610 \\ \hline
  (9,4)& 0.0600   & 0.0850 \\ \hline
\end{tabular}
\end{table}

\begin{table}[!]
\centering
\caption{Voltage magnitudes and phase angles at equilibrium A}
\begin{tabular}{|l|l|l|l|l|}
\hline
Bus & $V_i^s=V_i^0$ (p.u.) & $\theta_i^s$ (degee) \\ \hline
  1& 1       & 0   \\ \hline
  2& 1       & 5.1802   \\ \hline
  3& 1       & 5.5607  \\ \hline
  4& 0.9780  & 0.0791   \\ \hline
  5& 0.9542  & -0.4604   \\ \hline
  6& 0.9932  & 4.9809   \\ \hline
  7& 0.9818 & 3.9652   \\ \hline
  8& 0.9869  & 3.9639   \\ \hline
  9& 0.9673  & 0.7374   \\ \hline
\end{tabular}
\end{table}


\newpage

	
	\bibliographystyle{ieeetr}
	\bibliography{references}

\end{document}